\documentclass[aps,preprint,nofootinbib,%
superscriptaddress,amsmath,amssymb,showpacs]{revtex4}
\usepackage{epsfig}
\usepackage{bm}
\begin{document}
%
\preprint{INFNCA-TH0303}
\title{Antineutrinos from Earth: A reference model and its
       uncertainties}
\author{Fabio Mantovani}
\email{mantovani@fe.infn.it}
\affiliation{Dipartimento di Scienze della Terra, Universit\`a di Siena,
I-53100 Siena, Italy}
\affiliation{Centro di GeoTecnologie CGT,I-52027 San Giovanni Valdarno, Italy}
\affiliation{Istituto Nazionale di Fisica Nucleare, Sezione di Ferrara, 
I-44100 Ferrara, Italy} 
\author{Luigi Carmignani}
\email{luigi.carmignani@unisi.it}
\affiliation{Dipartimento di Scienze della Terra, Universit\`a di Siena,
I-53100 Siena, Italy}
\affiliation{Centro di GeoTecnologie CGT,I-52027 San Giovanni Valdarno, Italy}
\author{Gianni Fiorentini}
\email{fiorenti@fe.infn.it}
\affiliation{Dipartimento di Fisica, Universit\`a di Ferrara, I-44100
Ferrara, Italy}
\affiliation{Istituto Nazionale di Fisica Nucleare, Sezione di Ferrara, 
I-44100 Ferrara, Italy} 
\author{Marcello Lissia}
\email{marcello.lissia@ca.infn.it}
\affiliation{Istituto Nazionale di Fisica Nucleare, Sezione di Cagliari, 
             I-09042 Monserrato (CA), Italy}
\affiliation{Dipartimento di Fisica, Universit\`a di Cagliari,
             I-09042 Monserrato (CA), Italy}
%

\date{August 31, 2003; revised November 25, 2003}
\begin{abstract}
We predict geoneutrino fluxes in a reference model based on a detailed
description of Earth's crust and mantle and using the best available
information on the abundances of uranium, thorium, and potassium inside
Earth's layers. We estimate the uncertainties of fluxes corresponding to
the uncertainties of the element abundances. In addition to distance
integrated fluxes, we also provide the differential fluxes as a function
of distance from several sites of experimental interest. Event
yields at several locations are estimated and their dependence on the
neutrino oscillation parameters is discussed. At Kamioka we predict
$N(\text{U}+\text{Th})=35 \pm 6$ events for $10^{32}$ proton yr 
and 100\% efficiency assuming
$\sin^2(2\theta)=0.863$ and $\delta m^2 = 7.3  \times
10^{-5}$~eV$^2$. The maximal prediction is 55 events, obtained in a model
with fully radiogenic production of the terrestrial heat flow.
\end{abstract}
\maketitle
\section{\label{sec:intro}Introduction}

By looking at antineutrinos from reactors, KamLAND~\cite{Eguchi:2002dm} 
has confirmed  
the oscillation phenomenon previously discovered by SNO~\cite{Ahmad:2002jz}
with solar 
neutrinos and has provided crucial information on the oscillation 
parameters.
Putting together the results of solar and terrestrial experiments,
the best fit is obtained at $\delta m^2 = 7.3\times 10^{-5}$~eV$^2$ and
$\sin^2 (2\theta)=0.863$~\cite{Fogli:2002au}.  
Since we know their fate from production to detection,
neutrinos can now be used as physical probes.

Furthermore, the detector is so pure  and  the sensitivity   
is so high that  KamLAND will be capable 
of  studying  geoenutrinos, the antineutrinos originating  from  Earth's 
natural radioactivity.
Indeed, from a fit to the experimental data the KamLAND Collaboration 
reported four events associated with $^{238}$U and five with $^{232}$Th decay 
chains. This result, obtained from an exposure of just 
162~ton~yr, provides the first
insight into the radiogenic component of the terrestrial heat.
KamLAND has thus opened a new window for studying Earth's interior
and one expects more precise results in the near future from
KamLAND and other detectors which are presently in preparation.

The argument of geoneutrinos was introduced by Eder~\cite{Eder} in the
1960's and it was extensively reviewed by
Krauss~{\it et al.}~\cite{Krauss:zn} in
the 1980's.
Raghavan {\it et al.}~\cite{Raghavan:1997gw} 
and Rothschild {\it et al.}~\cite{Rothschild:1997dd}
\footnote{We shall always refer to the version available as arXiv:nucl-ex/9710001.} 
remarked on the potential of KamLAND and Borexino for geoneutrino observations. 
Fiorentini {\it et al.}~\cite{Fiorentini:2002bp,Fiorentini:2003ww,Fiorentini:2003pq} 
discussed the relevance of geoneutrinos
for determining the radiogenic contribution to Earth's heat flow and their 
potential for improving our knowledge of oscillation parameters, see also
Ref.~\cite{Nunokawa:2003dd}.

In preparation to the data which will become available in the near future,
we present a systematic discussion of geoneutrinos, which incorporates the
best geological and geochemical information on their sources and outlines
the main uncertainties, so as to understand what can be gained from the study
of geoneutrinos concerning both Earth's interior and neutrino properties.
With this spirit, we shall consider the following points.

(i) We provide a reference model that incorporates 
  the best available knowledge for the distribution of U, Th, and 
  K in Earth's interior.

(ii) Within this model we 
  predict neutrino fluxes and signals for detectors at different 
  positions on Earth.

(iii) We estimate uncertainties of neutrino fluxes and signals corresponding to 
uncertainties of the U, Th, and K distributions.

\section{The reference model: element distributions  and geoneutrino fluxes}

A global look at Earth's interior is useful before entering
a detailed
discussion on the element distributions.
 The amount of information which we (assume
to) have on Earth's interior is somehow surprising, 
if one considers that the
deepest hole which has ever been dug is probably only ten kilometers
deep, a mere dent in planetary terms.

The outer layer is the relatively thin crust, divided in two types,
continental crust (CC) and oceanic crust (OC). The former averages 38~km
in thickness, varying around the globe from 20 to 70~km, and it is
made primarily of light elements such as potassium, sodium,
silicon, calcium, and aluminium silicates. The oceanic crust is much
thinner, from about 6 to 8~km. 

Inside this crustal skin is Earth's
mantle which is 2900~km deep overall. Largely made up of iron and
magnesium silicates, the mantle as a whole accounts for about
68\% of Earth's mass. 
One distingushes the upper 
mantle~\footnote{We shall define the upper mantle as the
shallow mantle plus the transition region, i.e., the region
below the crust down to 677~km~\cite{PREM}.} 
(UM) from the lower mantle (LM),
however, the seismical discontinuities between the two parts do not 
necessarily divide the mantle into layers.
The main questions about the mantle are does it move as a single
layer or as multiple layers? Is it homogeneous in composition or
heterogeneous? How does it convect? These questions sound
simple, but the answers are complex, possibly leading to more
questions, see Ref.~\cite{grossman}.

Inside the mantle is Earth's
core, which accounts for about 32\% of Earth's mass.
Based on comparison with the behavior of iron at high pressures
and temperatures in laboratory experiments, on the seismic properties
of the core, and on the fact that iron is the only sufficiently 
abundant heavy element in the universe, the core is generally believed
to be made primarily of iron with small amounts of nickel and other
elements. Over thirty years ago, however, it was suggested that a
significant amount of potassium could be hidden in Earth's
core, thus providing a
large fraction of the terrestrial heat flow through $^{40}$K decay.
This
controversial possibility has been revived recently in Ref.~\cite{rama}.

Uranium, thorium, and potassium are lithophile elements, which
accumulate in the continental crust. Their abundance in the
mantle is much smaller, however, the total amounts are comparable
with those in the crust, due to the much larger mantle mass.
The core is generally believed to contain negligible amounts
of these elements.

A global description of the present crust-plus-mantle system is 
provided by the bulk silicate earth (BSE) model, 
a reconstruction of the primordial  mantle of Earth, subsequent to 
the core separation and  prior to crust differentiation, based on
geochemical arguments.
In the BSE model the uranium abundance~\footnote{We 
shall always refer to element abundances in mass  and we remind 
the reader that
the natural isotopic composition is  $^{238}$U/U = 0.993,
$^{232}$Th/Th = 1 and $^{40}$K/K = $1.2 \times 10^{-4} $.} 
is $a_{\text{BSE}}(\text{U})= 2 \times 10^{-8}$, and one has 
Th/U $\equiv a(\text{Th})/a(\text{U}) = 3.9 $
and
K/U $\equiv a(\text{K})/a(\text{U}) = 1.14 \times 10^4$, where the 
quoted values are averages between different estimates, 
all consistent with each other to the level of 10\% or 
better, see Table~\ref{table0}.
In the BSE model the total masses 
of uranium, thorium and potassium are thus 
$M(\text{U})=0.81\times 10^{17}$~kg, 
$M(\text{Th})=3.16
\times 10^{17}$~kg, and $M(\text{K}) = 0.49\times 10^{21}$~kg. 

The equation relating masses and heat production is
\begin{equation}
\label{eq:heat}
H=9.5 M(\text{U}) + 2.7 M(\text{Th}) + 3.6 M(\text{K})\quad ,
\end{equation}
where $H$ is in TW, $M(\text{U})$ and  $M(\text{Th})$
are in units of $10^{17}$~kg, and $M(\text{K})$  in units 
of $10^{21}$~kg.

In the BSE model, the contributed heat production rates  are
$H(\text{U})= 7.6$~TW, 
$H(\text{Th})=8.5$~TW, and $H(\text{K})=1.8$~TW, 
for a total of about one half of the observed terrestrial heat flow
($H_E\approx 40$~TW).

\subsection{Uranium, thorium, and potassium distributions}

Our aim is to build a reference model (labeled as ``ref''), 
which incorporates the best available 
knowledge of U, Th and K distributions inside Earth.
Concerning  Earth's crust, we distinguish oceans and seawater, 
the continental 
crust, subdivided into 
three sublayers (upper, middle, and lower), sediments and oceanic crust. 
All these layers have been mapped in Ref.~\cite{mappa}, 
which provides values of density and depth over the
globe on a grid with $2^{\circ}$ steps.   
We distinguish next the upper mantle (extending down to about 600 km), 
the lower mantle (down to about 2900~km), and the core, and use the 
preliminary reference earth model (PREM)~\cite{PREM} 
for the values of the density at each depth, 
assuming spherical symmetry.

For each component, one has to adopt a value for the abundances
$a(\text{U})$, $a(\text{Th})$, and $a(\text{K})$. In the literature
of the last twenty years
one can find  many estimates of abundances for the
various components of the crust (OC, upper CC, lower CC, \ldots),
generally
without an error value (see 
Tables~\ref{tab:abundancesU}, \ref{tab:abundancesTh}, and 
\ref{tab:abundancesK}), two classical reviews
being in Refs.~\cite{taylor85,wedepohl} and  a most useful
source being provided by the GERM Reservoir database~\cite{GERM}.

For the upper mantle we are aware of several 
estimates by Jochum {\it et al.}~\cite{jochum}, White~\cite{white},
O'Nions and McKenzie~\cite{onions}, Hofmann~\cite{hofmann}, and
Zartman and Haines~\cite{zartman}.
In this respect
data obtained from material emerged from unknown depths are 
assumed to be representative of the average composition down 
to about 600 km.

For each (sub)layer of the crust 
and for the upper mantle, we adopt as reference value for the uranium
abundance 
 $a^{\text{ref}}(\text{U}) $
the average of the values reported in 
Tables~\ref{tab:abundancesU}, \ref{tab:abundancesTh}, and 
\ref{tab:abundancesK}. 
Concerning Th and K, we observe that the abundance ratios  with 
respect to uranium
are  much more consistent among different authors than the corresponding 
absolute abundances. We shall thus 
take the average of ratios and from these construct the reference abundances for 
thorium and potassium: 
\begin{equation}
a^{\text{ref}}(\text{Th}) = \langle \text{Th/U} \rangle \; a^{\text{ref}}(\text{U})
\quad\quad \text{and}\quad \quad  
a^{\text{ref}}(\text{K}) = \langle \text{K/U} \rangle \; a^{\text{ref}}(\text{U})  \quad .
\end{equation}
For the lower mantle, where no observational data are available, we resort 
to the BSE model, which ---  we recall --- describes  
the present crust-plus-mantle system based on geochemical arguments.  

The mass of each element ($X$ = U, 
Th, K) in the lower mantle $M_{\text{LM}}(X)$ is thus obtained by subtracting 
from the BSE estimate the mass calculated for the crust and upper mantle:
\begin{equation}
\label{BSEconstrain}
M_{\text{LM}}(X)= M_{\text{BSE}}(X) - M_{\text{CC}}(X) 
- M_{\text{OC}}(X) - M_{\text{UM}}(X) \quad .
\end{equation}
Reference abundances for the lower mantle are then obtained by
dividing these values by its mass $M_{\text{LM}}= 2.9\times 10^{24} $~kg.   
According  to geochemical arguments, negligible amounts of U,
Th and K should be present in the core.

The resulting choice of input values for the reference model is collected  
in Tables~\ref{tab:abundancesU}, \ref{tab:abundancesTh},
and \ref{tab:abundancesK}.
Concerning this reference model, we remark the following points. 

(i)
The uranium mass in the crust $M_{c}(\text{U}) = 0.35\times 10^{17}$~kg 
is mainly  concentrated in the continental part. The oceanic crust contributes  
as little as $0.005\times 10^{17}$~ kg, since its impoverished  by a factor of  
20 and it is much thinner than the continental crust.

(ii) 
The estimated uranium mass in the upper mantle is about one sixth of that in the 
crust, whereas the lower  mantle contains about as much uranium 
as the crust.

(iii)
Note that in this refererence model, constructed  so as to satifisy the 
BSE constraint (\ref{BSEconstrain}), mantle
depletion (with respect to BSE) extends to the lower mantle.

(iv)
Similar considerations hold for thorium and potassium.

\subsection{The reference fluxes}

For each element $X$ the produced~\footnote{The produced fluxes are
calculated ignoring oscillations, which will be discussed later.}
antineutrino fluxes at position 
$\vec{r}$ are defined as
\begin{equation}
\label{eq:total}
\Phi_{X}(\vec{r}) = \frac{n_X}{4\pi \mu_{X}\tau_X} 
   \int_{V_{\oplus}}    d\vec{r'}
   \frac{\rho(\vec{r'})a_{X}(\vec{r'})}{|\vec{r}-\vec{r'}|^2} \quad ,
\end{equation}
where  $\tau$ is the lifetime, $\mu$ is the atom mass, $n$ is the number 
of antineutrinos per decay chain, the integral is over the  
volume of the earth,    
$\rho$ is the local density, and $a_{X}$ is the abundance of the element
$X$. 
We have evaluated   
the produced fluxes at several sites on the globe within the
reference model ($a=a^{\text{ref}}$).  
We concentrate here on a few locations of specific interest, see 
Tables~\ref{tab:fluxU}, \ref{tab:fluxTh}, and \ref{tab:fluxK}.

(i) for the Kamioka mine, where the  KamLAND detector is 
in operation, we predict an uranium flux 
$\Phi_U = 3.7\times 10^6 \text{ cm}^{-2}\text{ s}^{-1}$, 
a  comparable flux from thorium and a fourfold flux from potassium. 
Within the reference model, about 3/4 of the flux is generated from 
material in the crust and the rest mainly from the lower mantle.

(ii) At Gran Sasso laboratory, where Borexino~\cite{borexino} 
is in preparation, we  predict an uranium flux  
$\Phi_{\text{U}} = 4.2\times 10^6 \text{ cm}^{-2}\text{ s}^{-1}$, 
this larger flux arising from a bigger contribution of the surrounding 
continental crust. Thorium and potassium fluxes are correspondingly 
rescaled. 

(iii) At the top of Himalaya, a place chosen so that the crust 
contribution is maximal,  we find the maximum  uranium flux
 $\Phi_{\text{U}} = 6.7\times 10^6 \text{ cm}^{-2}\text{ s}^{-1}$.
The crust  contribution exceeds 90\%.

(iv) On the Hawaii, a site which minimizes the crust contribution, 
 we find   $\Phi_{\text{U}} = 1.3\times 10^6 \text{ cm}^{-2}\text{ s}^{-1}$,
originated mainly from the mantle.

These computed reference fluxes are generally larger than those of 
Rothschild~\cite{Rothschild:1997dd}, by a
factor of order 30--50 \%. This arises from several differences in 
the approaches.

(i)
We have used a more recent and detailed map of Earth's crust: 
the grid is denser and several layers are distinguished.

(ii)
We have a more detailed model for the mantle, correponding to 
the PREM density profile.

(iii) 
Most important, our reference values for the abundances in the continental 
crust are larger than that used in Ref.~\cite{Rothschild:1997dd}. 
As an example, Rothschild~{\it et al.} use for the continental crust 
$a_{\text{CC}}(\text{U}) = 0.91$~ppm  from a classical review paper
of 1985~\cite{taylor85}. Our reference  model, when averaged over
the different sublayers, yields  $a_{\text{CC}}(\text{U}) = 1.5$~ppm. 
This larger value arises from taking into account recent data, which are  
all  higher than those quoted in Ref.~\cite{taylor85}.

The produced fluxes are computed ignoring the effect of oscillations,
which depends on the distance $R$ between the detector and the source.
For taking into account this effect, and also in view of 
understanding which portion of Earth can be accessed with a 
geoneutrino detector, it is useful to introduce  quantities which contain
more detailed information.

The differential fluxes $f(R)$ are obtained by grouping together all the 
sources which lie at  the same distance $R$ from the detector
\begin{equation}
\label{eq:diff}
 f_{X}(R) =  \frac{n_X}{4\pi \mu_{X}\tau_X} 
 \int_{V_{\oplus}} d\vec{r'}
   \frac{\rho(\vec{r'})a_{X}(\vec{r'})}{|\vec{r}-\vec{r'}|^2}
   \delta(R-|\vec{r}-\vec{r'}|) \quad .
\end{equation}
Note that $ f_{X}(R)$ actually depends also on the detector 
position $\vec{r}$ and just 
for semplicity of notation we drop this variable.

The cumulated fluxes $\phi(R)$ are defined as
\begin{equation}
\label{eq:cumu}
\phi_{X}(R) = \int_0^R dR'  f_{X}(R') \quad . 
\end{equation}
They represent the cumulative effect of all sources within a distance
$R$ from the detector: the total produced fluxes of Eq.~(\ref{eq:total})
are clearly $\Phi_X = \phi_{X}(2 R_{\oplus}) $.

\begin{figure}[p]
\psfig{figure=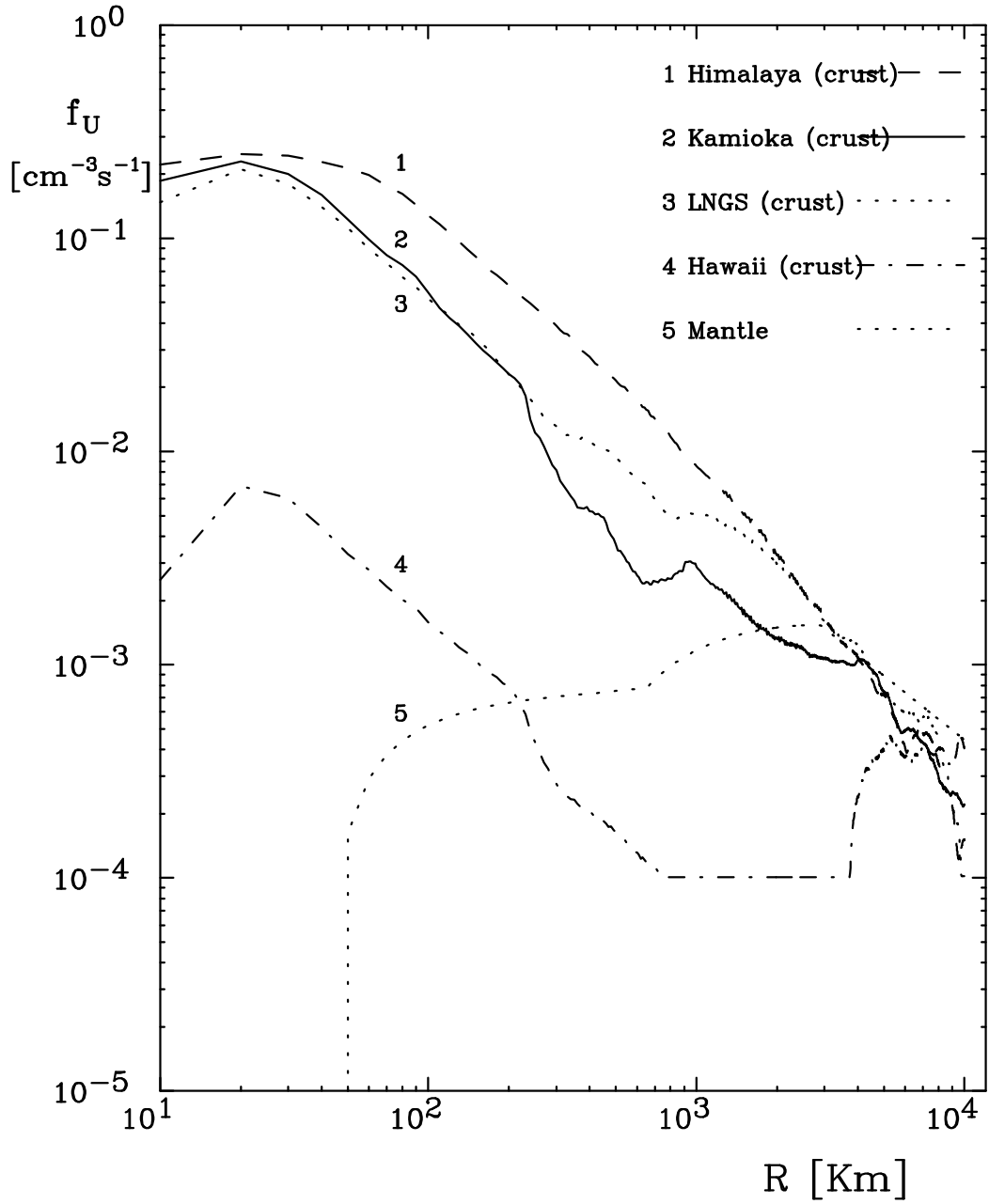,%
bbllx=80pt,bblly=80pt,bburx=560pt,bbury=730pt,%
height=20cm}  
\caption{
Differential produced flux from uranium as a function of the 
distance $R$ from the detector.
  \label{fig:diff}
}
\end{figure}

\begin{figure}[p]
\psfig{figure=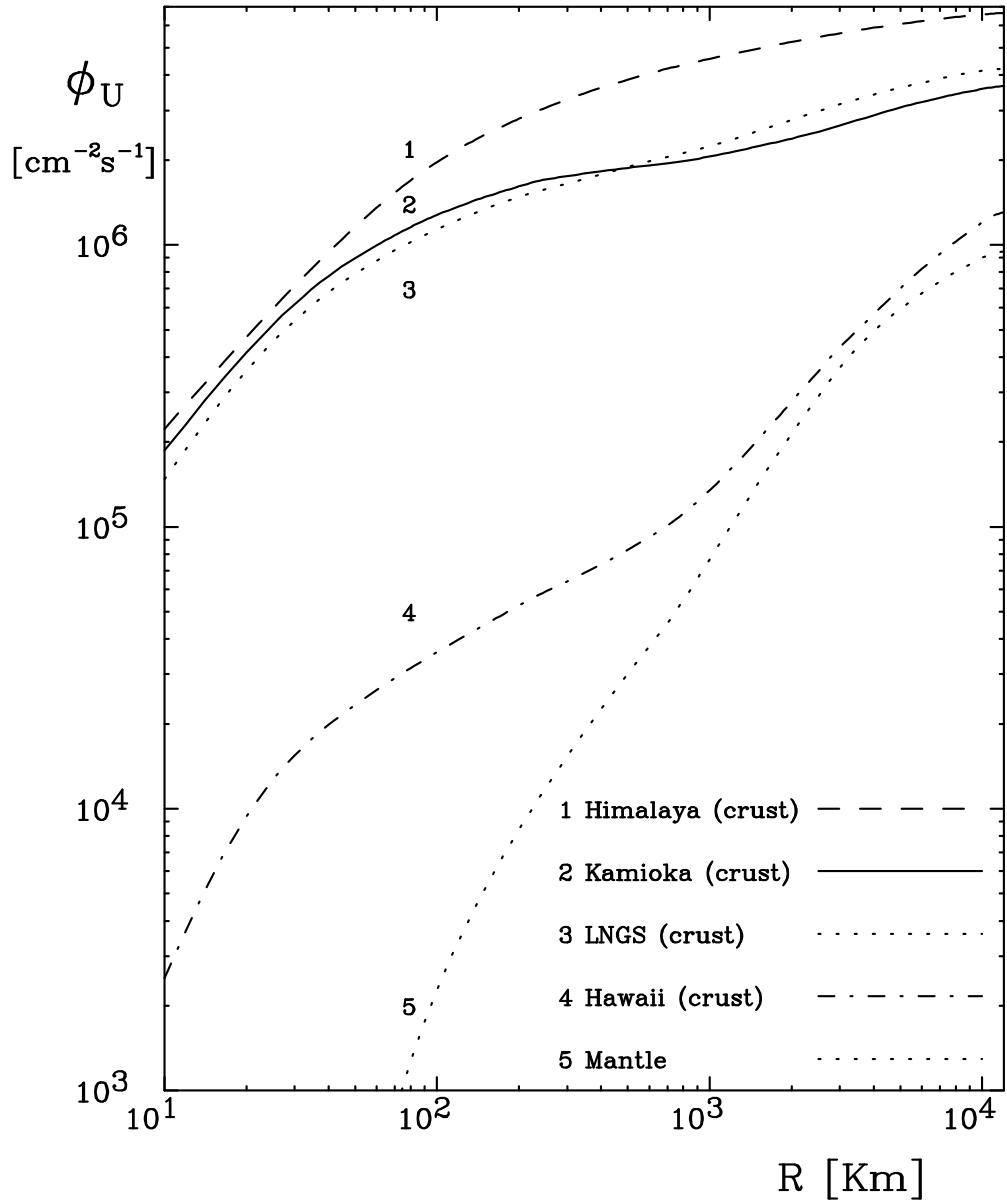,%
bbllx=70pt,bblly=70pt,bburx=550pt,bbury=750pt,%
height=20cm}
\caption{
Cumulated produced flux from uranium as a function of the 
distance $R$ from the detector.
  \label{fig:cumu}
}
\end{figure}

These quantities are plotted in Figs.~\ref{fig:diff}
and \ref{fig:cumu} for the
four sites (we only show the uranium contribution, the shapes of the other
contributions being similar). 
We remark that we have been using an average density approximation,
which presumably breaks down near the detector, where one should
resort to a detailed geological study of the surroundings.
From Fig.~\ref{fig:cumu} one sees that in our model 
the region within 30~km from  Kamioka or Gran Sasso originates
about 15\% of the total produced flux. 
Concerning the region where most of the flux is generated, one sees
again from  Fig.~\ref{fig:cumu} that 50\% of the produced flux
originated within 400~km (800~km) from Kamioka (Gran Sasso). 

In Tables~\ref{tab:Kami}, \ref{tab:GrSa} and \ref{tab:mantle} we present
the numerical values of the contribution to $f_X(R)$ from the crust 
at Kamioka and Gran Sasso and that from the mantle (the assumed 
spherical symmetry of the mantle implies the same contribution at
any site). These data will be useful for a detailed analysis of future
experiments devoted to the study of geoneutrinos, in order to take
into account the distance dependence of the survival probability.

\section{The uncertainties of the reference model}

The fluxes of the reference model correspond to the best available knowledge 
about the crust and 
the interior of Earth, as derived from observational data and 
geochemical information on the global 
properties. An estimate of the uncertainties of the predicted fluxes 
is clearly useful.

Since the abundance ratios look relatively well determined, we 
concentrate on the uncertainties of the uranium abundances in 
the different layers and propagate them to the other elements.
For the reference model, we 
have $M_{\text{CC}}(\text{U})= 0.345 \times 10^{17}$~kg, 
$M_{\text{OC}}(\text{U}) = 0.005  \times 10^{17}$~kg, 
the total mass of CC being  $M_{\text{CC}}= 2.234 \times 10^{22}$~kg. 
According to our model, 
the average uranium abundance in the CC is thus 
$a_{\text{CC}}(\text{U})= 1.54\times 10^{-6}$.
We determine lower and upper limits  by observing that the range of 
estimated  uranium abundances  
is between  $0.91\times 10^{-6}$~\cite{taylor85} and 
 $1.8\times 10^{-6}$~\cite{shaw86}

\begin{center}
\begin{tabular}{lccc}
   low:  & $a_{\text{CC}}(\text{U})=0.9  \times 10^{-6}$; 
         & $a_{\text{CC}}(\text{Th})=3.7 \times 10^{-6}$; 
         & $a_{\text{CC}}(\text{K})=0.94 \times 10^{-2}$ \quad ,\\ 
   high: & $a_{\text{CC}}(\text{U})=1.8  \times 10^{-6}$; 
         & $a_{\text{CC}}(\text{Th})=7.6 \times 10^{-6}$; 
         & $a_{\text{CC}}(\text{K})=1.97 \times 10^{-2}$ \quad .\\
\end{tabular}
\end{center}         
We remark that there is an overall uncertainty of a factor 2 concerning the 
total amount of radioactive materials in the crust.

For the upper mantle, we take as extrema the 
two values known to us~\cite{jochum,zartman} 
for uranium and we deduce thorium and potassium by rescaling

\begin{center}
\begin{tabular}{lccc}
   low:  & $a_{\text{UM}}(\text{U})=5  \times 10^{-9}$; 
         & $a_{\text{UM}}(\text{Th})=13 \times 10^{-9}$; 
         & $a_{\text{UM}}(\text{K})=\phantom{9.}6 \times 10^{-5}$ \quad ,\\ 
   high: & $a_{\text{UM}}(\text{U})=8  \times 10^{-9}$; 
         & $a_{\text{UM}}(\text{Th})=21 \times 10^{-9}$; 
         & $a_{\text{UM}}(\text{K})=9.6 \times 10^{-5}$ \quad .\\
\end{tabular}
\end{center}
Such a small uncertainty  is  perhaps optimistic, however, it is not 
influential  for the future discussion  in view of the relatively small 
amounts contained in the upper mantle.

We remind the reader that no observational information is 
available for the lower mantle.
For building a minimal model, we assume that the mantle is fully mixed
and use for the whole  mantle the lowest values estimated 
from samples coming from the upper mantle.

A maximal model can be obtained by assuming that the terrestrial
heat is fully accounted by radiogenic production. 
This can be otained 
by keeping the  BSE abundance ratios fixed and rescaling the total masses to 
$M(\text{U})=1.67 \times 10^{17}$~kg, $M(\text{Th}) = 6.5 \times 10^{17}$~kg,
and $M(\text{K})= 1.9\times 10^{21}$~kg.~\footnote{Clearly  
this model does not satisfy the BSE
constraint on the total U, Th, and K masses in the Earth.}
A natural implementation is obtained by
choosing for the crust and upper mantle the highest observational 
estimates and placing the remaining mass in the lower mantle.

All this  leads to
\begin{center}
\begin{tabular}{lccc}
   low:  & $a_{\text{LM}}(U)=5  \times 10^{-9}$; 
         & $a_{LM}(\text{Th})=13 \times 10^{-9}$; 
         & $a_{LM}(\text{K})=\phantom{45.}6 \times 10^{-5}$ \quad ,\\ 
   high: & $a_{LM}(\text{U})=40  \times 10^{-9}$; 
         & $a_{LM}(\text{Th})=156 \times 10^{-9}$; 
         & $a_{LM}(\text{K})=45.6 \times 10^{-5}$ \quad .\\
\end{tabular}
\end{center}
The corresponding low and high estimates of the produced fluxes  
are also shown in Tables~\ref{tab:fluxU}, \ref{tab:fluxTh},
and \ref{tab:fluxK} for a 
few locations. 

In view of assigning an uncertainty to the fluxes of the reference 
model one can take two different approaches.

(a) A conservative estimate: the error assigned to the
reference value is half 
of the difference between the high and low estimates
$\Delta \Phi_{\text{cons}} =  ( \Phi_{\text{high}}-\Phi_{\text{low}})/2$.

(b) A statistical estimate: one assumes that
the full range of calculated fluxes represents
a $\pm 3 \sigma$ interval.~\footnote{
If unhappy with this conventional
assumption, the reader can rescale $\sigma$.} In this
way one obtain a conventional $1\sigma $ error
 $\Delta \Phi =  ( \Phi_{\text{high}}-\Phi_{\text{low}})/6$.

The  relative uncertainties of the fluxes
are reported in Table~\ref{tab:error}. 
They are the same (and fully correlated) for all elements, the
1-$\sigma$ 
error being  about 15\%, at Kamioka and Gran Sasso. At Hawaii, 
where the mantle contribution is
dominant, the error is much larger, as a consequence of the large
uncertainties of the lower mantle's composition.

When using these errors, one has to remark that uncertainties associated 
with abundances in the crust and in the upper mantle are deduced
from the spread of observational data, whereas the estimates for the 
lower mantle, which cannot be accessed by observations,  completely rely on 
theoretical arguments. In addition, one should also take into account
the detailed geological
structure around the detector for more precise flux estimates.

\section{From fluxes to signals and detectors}

Geoneutrinos can be detected by means of inverse beta reactions
\begin{equation}
\label{inversebeta}
 \bar{\nu}_e + (Z,A) \to e^+ + (Z-1,A) \quad ,
\end{equation}
where the positron kinetic energy $T$ is related to the 
antineutrino energy $E$ by 
       $T= E -E_0$, 
with   
$E_0 = m_{Z-1} + m_e- m_{Z}$.~\footnote{A 
frequently used variable is the visible energy $E_{\text{vis}} = T + 2 m_e$ which
is the energy released in the slowing down and subsequent annihilation
of the positron.}
The  differential event yield  as a function of $T$ is given by
\begin{equation}
\label{eq:diffspec}
\frac{dN}{dT} = N_Z t \sigma(E) \sum_X w_X(E)\int_0^{2 R_{\oplus}} dR f_X(R) 
P_{ee} (E,R) \quad ,
\end{equation}
where $N_Z t$ is the exposure 
(number of target nuclei times the live time), $\sigma(E)$ is the 
cross section of reaction~(\ref{inversebeta}),
$T= E -E_0$ and 
the integral is over the distance $R$ from the detector.

The survival probability of $\bar{\nu}_e $ produced
at distance $R$ with energy $E$ is
\begin{equation}
\label{eq:survprob}
P_{ee}(E,R) = 1 - \sin^2(2\theta) \sin^2\left(\frac{\delta m^2 R}{4E}  \right)
 \quad .
\end{equation}
For each element,
the  differential produced flux 
$f_X (R)$ is defined 
in Eq.~(\ref{eq:diff}),
$w_X(E)$ is the energy spectrum of the $\bar{\nu}_e $ from the decay
chain~\cite{behrens} of element $X$ and 
normalized to 1, $\int_0^{\infty} dE w_X (E) = 1 $. For
simplicity we neglect the finite energy resolution of the
detector and assume 100\% detection efficiency.

Another interesting observable is the total geoneutrino
yield
\begin{equation}
N = \int_0^{T_{\text{max}}} dT \frac{dN}{dT} \quad ,
\end{equation}
where $T_{\text{max}}$ is the maximal positron energy.

The classical approach to low energy antineutrino detection 
is by using  hydrogen compounds as target, by 
means of  $\bar{\nu}_e + p \to  e^+ + n $. 
Since $E_0 = m_n + m_e - m_p = 1.804$~MeV, this reaction is suitable for  
antineutrinos from  uranium and thorium progenies 
($E_{\text{max}} = 3.26$ and 2.25~MeV, respectively), 
whereas antineutrinos from potassium ($E_{\text{max}}=1.31$~MeV) are 
below threshold.

\subsection{Total yields}

We discuss first the total geoneutrino yield $ N$, which is 
experimentally more accessible than the differential spectrum.
In view of the structure of the survival probability, 
see Eq.~(\ref{eq:survprob}), 
it can be written as
\begin{equation}
\label{eq:chi}
N = N_{\text{no}} \left[ 1 -  \sin^2(2\theta)  \chi(\delta m^2) \right]
\quad ,
\end{equation}
where $ N_{\text{no}}$ is the yield for no oscillation.

\begin{figure}[p]
\psfig{figure=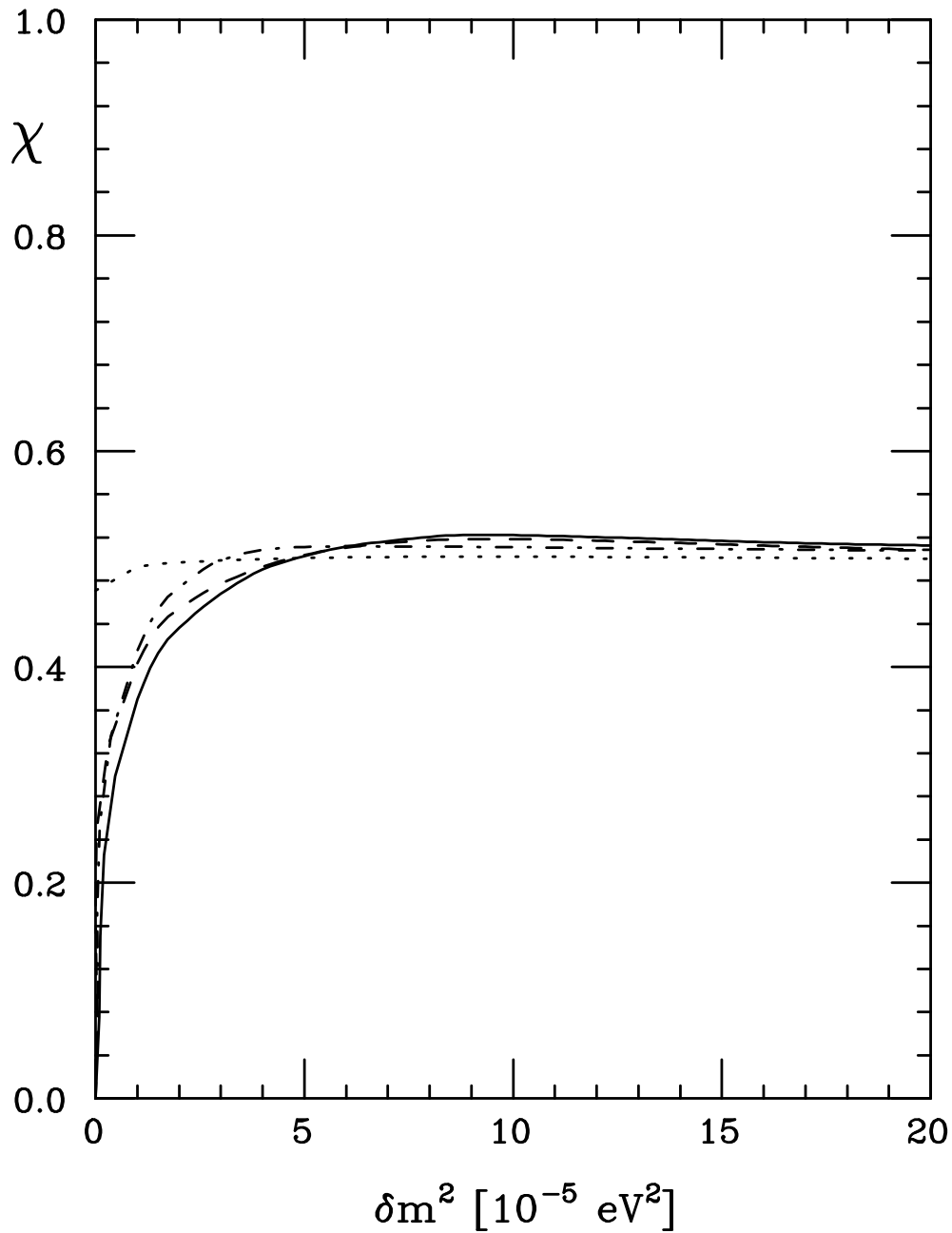,%
bbllx=80pt,bblly=70pt,bburx=560pt,bbury=720pt,%
height=20cm}  
\caption{
Dependence of the yield on on $\delta m^2$. 
The figure shows the
function $\chi = (N_{\text{no}}-N)/\left[(N_{\text{no}}\sin^2(2\theta))\right]$, 
see Eq.~(\ref{eq:chi}),
for four locations with 
$\delta m^2$ in units of
$10^{-5}$~eV$^2$.
Solid (dashed, dotted, dot-dashed) line applies to
Kamioka (LNGS, Hawaii, Himalaya). 
  \label{fig:chi}
}
\end{figure}

\begin{figure}[p]
\psfig{figure=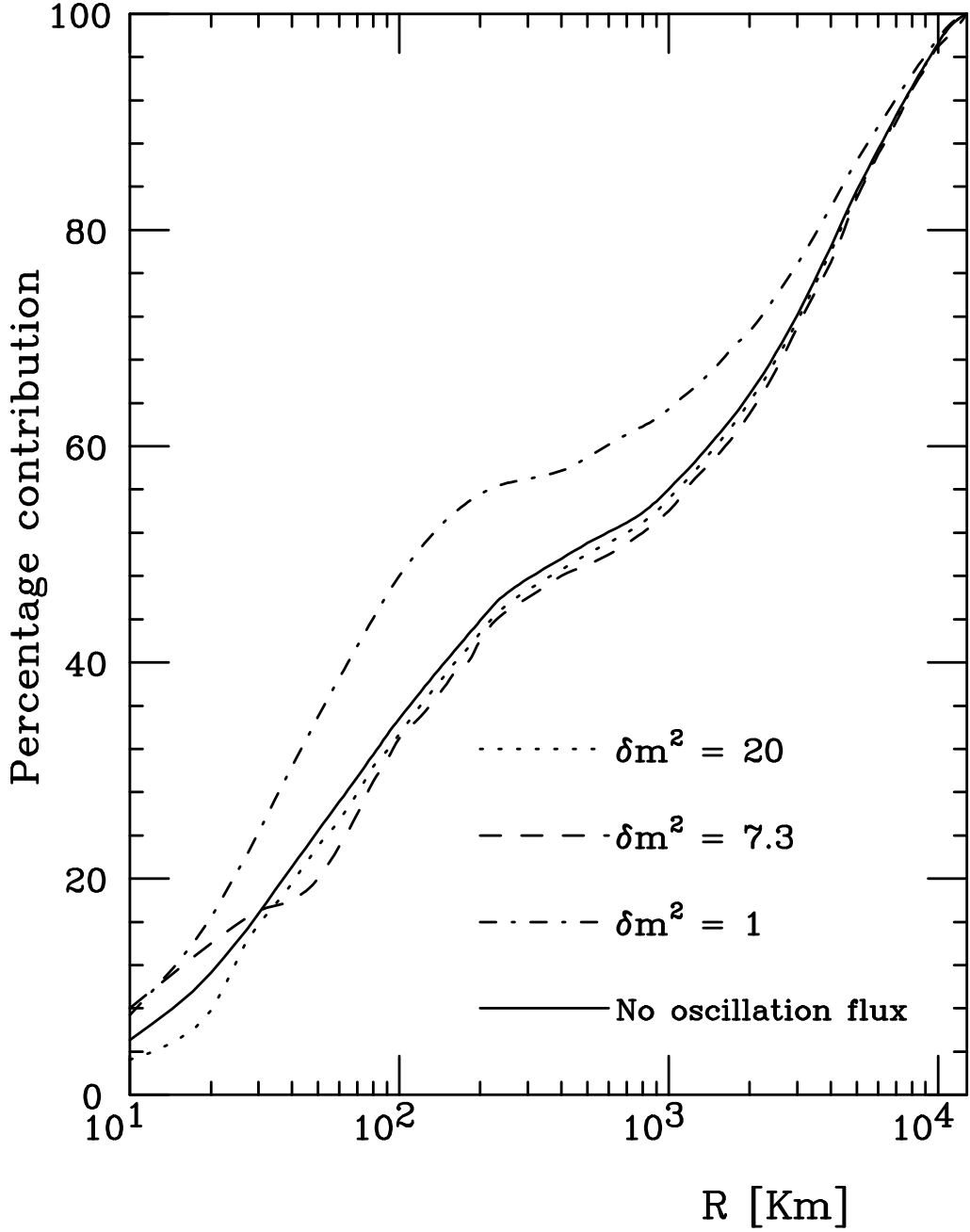,%
bbllx=60pt,bblly=100pt,bburx=560pt,bbury=710pt,%
height=19cm}  
\caption{
Contributed signal as a function of distance. The percentage
contribution to the event yield at Kamioka originating from sources
within $R$ is shown for the indicated values of $\delta m^2$ in units of
$10^{-5}$~eV$^2$ at fixed $\sin^2(2\theta)=0.863 $. The percentage contributed
neutrino flux without oscillation is also shown for comparison. 
  \label{fig:cumusign}
}
\end{figure}

The function $\chi$ depends on the uranium and thorium
distributions inside Rarth and on the detector position.
Obviously $\chi$ tends to 0 (1/2) for small (large) values
of $\delta m^2$.
We have computed  $\chi$ in the reference model for some
sites of interest, see Fig.~\ref{fig:chi}. 
At all locations and  for $\delta m^2 > 4\times 10^{-5}$~eV$^2$,
the function $\chi$ differs
from its asymptotic value by less than 2\%. 
Using the asymptotic value of the survival probability and
the best fit value of the mixing angle~\cite{Fogli:2002au},
one finds
\begin{equation}
N = N_{\text{no}} \left[ 1-0.5\sin^2(2\theta) \right] 
  = 0.57 N_{\text{no}} \quad .
\end{equation}
In Fig.~\ref{fig:cumusign} we show the relative contributions of
different distances to the total yield: for the most
interesting values of $\delta m^2$ the region within 30~km
from Kamioka contributes about 15\% of the total.
The no oscillation yield $N_{\text{no}}$ is determined in terms
of the total produced fluxes from uranium and 
thorium~\cite{Fiorentini:2003ww}
\begin{equation}
N_{\text{no}} = 13.2 \Phi_{\text{U}} + 4.0 \Phi_{\text{Th}} 
\end{equation}
for an exposure of $10^{32}$ proton yr
with
fluxes $\Phi$ in units of $10^6$~cm$^{-2}$~s$^{-1}$.

The no oscillation yields, calculated with the fluxes of
the reference model, are shown in  Table~\ref{tab:events}.
In the same table we also present the estimated 
$1 \sigma$ errors,
obtained by propagating those on the produced fluxes
(which are dominant over the other uncertainties
from cross sections, decay spectrum, etc.) and the
minimal and maximal predictions.

For the Kamioka site the prediction of the reference model
(normalized~\footnote{It
is useful to introduce a terrestrial neutrino unit (TNU)
for event rates, 
defined as one event per $10^{32}$ target nuclei per year,
or $3.171\text 10^{-40}$~s$^{-1}$ per target nucleus. This unit
is analogous to the solar neutrino unit (SNU)~\cite{Bahcall:ks}.}
to $10^{32}$ proton yr
and 100\% efficiency) is $N_{\text{no}} = 61$ in good agreement
with the ``best model'' 
of Refs.~\cite{Fiorentini:2002bp,Fiorentini:2003pq},
 $N_{\text{no}}= 67 $~TNU,
in between the values of  Ref.~\cite{Rothschild:1997dd},
 $N_{\text{no}}= 43 $~TNU,
and of model 1b of Ref.~\cite{Raghavan:1997gw}, 
 $N_{\text{no}}= 75 $~TNU.
An experimental value for $N_{\text{no}}= 156 $~TNU can be deduced from
the nine geoevents reported by KamLAND, assuming
$P_{ee} = 0.57$. 
All the above predictions are consistent
with the experimental result
within its statistical error 
(about 60\%~\cite{Fiorentini:2003ww}). 

The total yields predicted in our
reference model for a number of locations are presented
in Fig.~\ref{fig:yields}.
We remind the reader that geoneutrino fluxes are superimposed to the 
low-energy tail of antineutrinos from nuclear reactors, which can
provide in this respect an important background, as first pointed 
out by Lagage~\cite{lagage85}. 
This effect is clearly dependent on location and it has been extensively
discussed in Ref.~\cite{Fiorentini:2003pq}. In particular, the event
yield from reactors has been estimated as about 300 TNU (no oscillations) 
at Kamioka and about 70 TNU at Gran Sasso.

\subsection{Event spectra}
A more detailed information is contained in the event
spectrum $dN/dT$ and a relevant question is whether
the spectrum is deformed because of oscillations.
 From Eqs~(\ref{eq:diffspec}) and (\ref{eq:survprob}) 
the event distribution with energy
can be written as
\begin{equation}
\label{eq:defspectrum}
 \frac{dN}{dT}
= \left(\frac{dN}{dT}\right)_{\text{no}} 
   \left[ 1-\sin^2(2\theta) \psi(T,\delta m^2) \right]
    \quad ,
\end{equation}
where $T$ is the positron kinetic energy.

\begin{figure}[p]
\psfig{figure=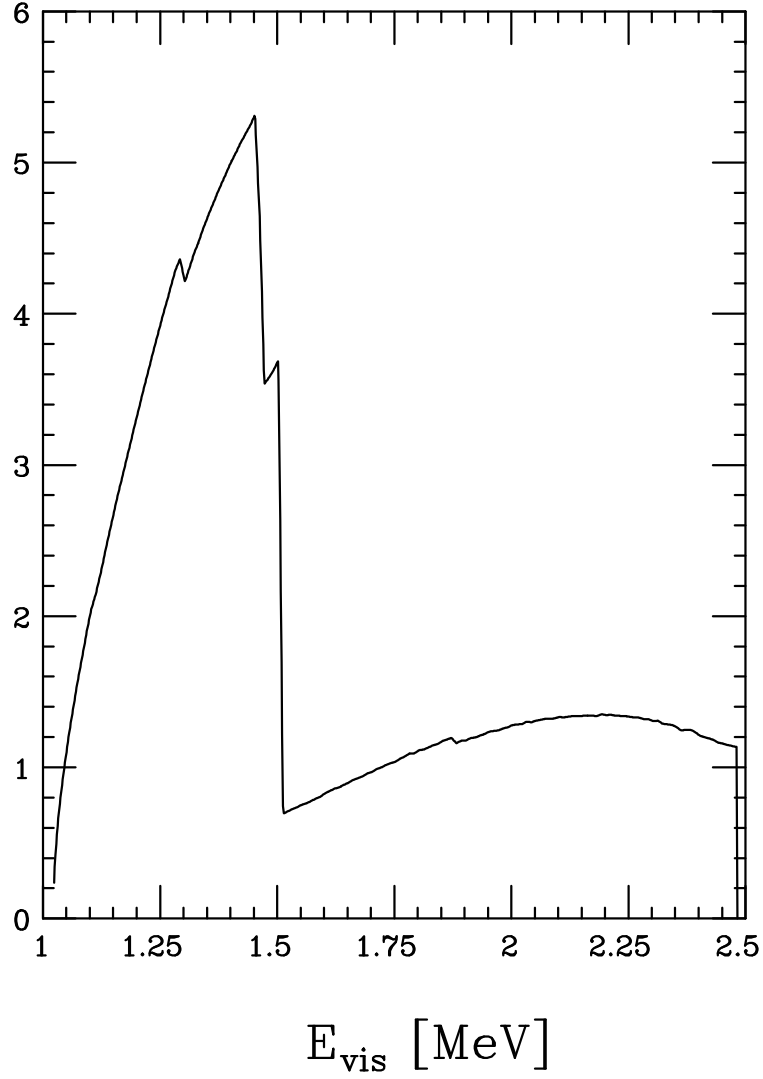,%
bbllx=70pt,bblly=70pt,bburx=560pt,bbury=720pt,%
height=16cm}  
\caption{
Event spectrum
as function of the 
visible energy $E_{\text{vis}}=T+2m_e$ in MeV.
The spectrum is calculated for the U/Th flux ratio expected at 
Kamioka with no oscillation 
and the normalization is arbitrary. 
  \label{fig:spectra}
}
\end{figure}

\begin{figure}[p]
\psfig{figure=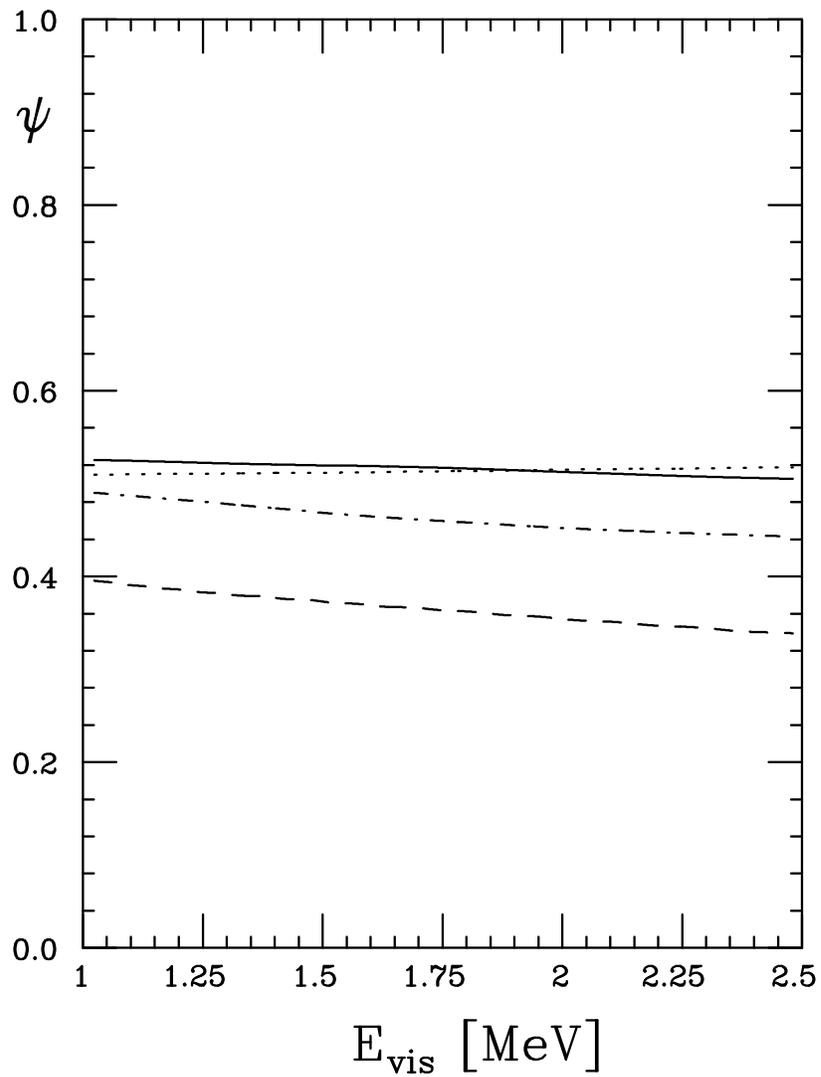,%
bbllx=60pt,bblly=75pt,bburx=555pt,bbury=710pt,%
height=16cm}  
\caption{
Spectrum deformation.
The function
$\psi$,
defined in Eq.~(\ref{eq:defspectrum}),
as function of the visible energy 
$E_{\text{vis}}=T+2m_e$ in MeV
for four  values of $\delta m^2$:
$1\times 10^{-5}$~eV$^2$ (dash line),
$3\times 10^{-5}$~eV$^2$ (dot-dash line),
$7.3 \times 10^{-5}$~eV$^2$ (solid line),
and $20\times 10^{-5}$~eV$^2$ (dot line).
  \label{fig:deformspectra}
}
\end{figure}

The no-oscillation spectrum $dN_{\text{no}}/dT$ is shown in
Fig.~\ref{fig:spectra} for Kamikande. 
The function  $\psi(T, \delta m^2)$ represents the modification to the
event spectrum due to oscillations. It  is
plotted for Kamioka for a few values of 
$\delta m^2 $ in Fig.~\ref{fig:deformspectra}.  
One sees that oscillations produce
a moderate distortion for the two smallest values of  $\delta m^2 $ and
the distortion is negligible for the largest values of $\delta m^2 $.

\section{Concluding remarks}

We summarize here the main points of this paper

(i) 
We have provided a reference model for the produced fluxes of geoneutrinos,
estimating its uncertainties in view of available data and geochemical 
inferences about U, Th, and K distribution in 
Earth's interior. When normalized to an exposure of 
 $10^{32}$ proton yr,
an averaged  survival 
probability $\langle P_{ee}\rangle  = 0.57$ and a  100\% 
detection efficiency, the  predicted events for KamLAND are
\begin{equation}
       N(\text{U})   = 28  \pm 4.7  \quad\quad , \quad\quad     
       N(\text{Th})  =  7  \pm 1.2 \quad . 
\end{equation}
Errors have been estimated so as 
correspond to $1\sigma$  confidence level and are (almost)
completely correlated:
\begin{equation}
  N(\text{U} + \text{Th}) = 35 \pm 6   \quad .
\end{equation}

(ii)
Concerning the estimated errors,
we remark that uncertainties associated 
with abundances in the crust and in the upper mantle are deduced
from the spread of observational data, whereas the estimates for the 
lower mantle, which cannot be accessed by observations,  completely rely on 
theoretical arguments. In addition, one should also take into account
the detailed geological
structure around the detector for more precise flux estimates.

(iii) 
We have also investigated extreme models, corresponding the the 
minimal and maximal amounts 
of U and Th  which could be present  on Earth. 
At KamLAND we predict
\begin{equation}
 N^{\text{low}}(\text{U} + \text{Th}) =  29 \quad\quad \text{and} \quad\quad
  N^{\text{high}}(\text{U} + \text{Th}) =  74 \quad .
\end{equation}
In these two  extreme models U, Th, and K, in the BSE proportions, 
produce a radiogenic heat $H_{\text{rad}} = 9$ and 40~TW, respectively. 
If experimental results come out close to the minimum, then  
uranium and thorium provide a 
minor contribution to Earth's energetics: either Earth's  heat flow is 
mainly non radiogenic or a 
significant amount of potassium has to be hidden  in Earth's interior. 
If  values  near to the maximal are found from experiments, 
then radiogenic contribution is the main supply of 
Earth's heat flow, and one can exclude models where significant 
amounts of potassium are hidden in Earth's core.      

\begin{figure}[p]
\psfig{figure=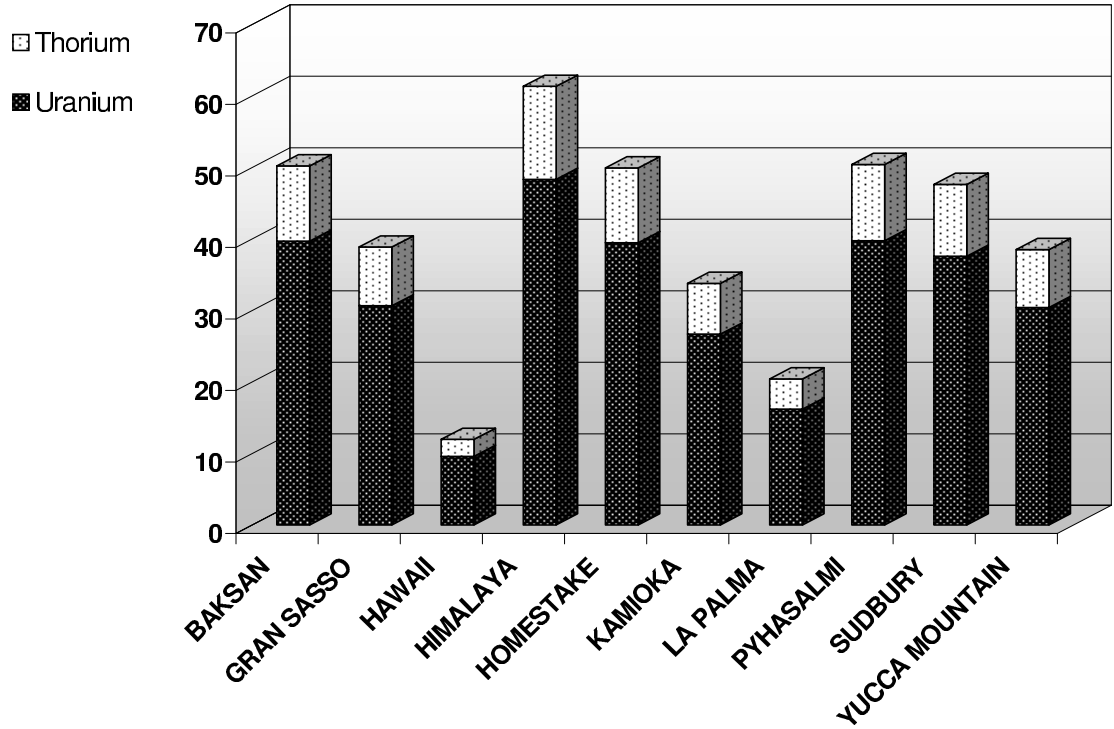,%
bbllx=74pt,bblly=60pt,bburx=520pt,bbury=780pt,%
height=16cm,angle=-90}
\caption{
Yields predicted in the reference model for
$10^{32}$ proton yr, 
100\% efficiency,
assuming the best fit oscillation parameters, 
$\delta m^2 = 7.3\times 10^{-5}$~eV$^2$ and
$\sin^2(2\theta) = 0.863$.
  \label{fig:yields}
}
\end{figure}

(iv) Predictions for detectors at several  locations are also given, 
see Table~\ref{tab:events} and Fig.~\ref{fig:yields}.
We remark that a 
detector located far from the continental crust could provide significant
information on the structure of the mantle, particularly when compared 
with data from  detectors  at sites  where (as in 
KamLAND and Borexino) the contribution of Earth's crust  is important.

\begin{acknowledgments}
We are grateful for useful comments and discussions to B.~Ricci,
L.~Beccaluva, T.~Lasserre, E.~Lisi, R.~Marcolongo, G.~Ottonello,
S.~Sch{\"o}nert, and R.~Vannucci.
This work was partially supported by MIUR (Ministero dell'Istruzione,
dell'Universit\`a e della Ricerca) under PRIN 2001
and  PRIN 2002. 
\end{acknowledgments}

\clearpage

\begin{table}
\caption[aaa]
{\label{table0}
Abundances in the bulk silicate Earth model.
}
\begin{ruledtabular}
\begin{tabular}{llll}
$a(\text{U})$ & Th/U & K/U & Remarks \\
\hline
$2.1\times 10^{-8}$ & 4.0  &  $1.14\times 10^{4}$& \cite{mcdonough92} \\
$2.3\times 10^{-8}$ &      &                     & \cite{wanke84} \\
$2.0\times 10^{-8}$ & 4.0  &  $1.27\times 10^{4} $& \cite{hofmann} \\
$1.8\times 10^{-8}$ & 3.6  &  $1.0 \times 10^{4} $& \cite{taylor85} \\
\hline
$2.0\times 10^{-8}$ & 3.9  &  $1.14\times 10^{4} $& average       \\
\end{tabular}
\end{ruledtabular}
\end{table}

\begin{table}
\caption[bbb]
{\label{tab:abundancesU}
Uranium abundances in Earth's interior.
}
\begin{ruledtabular}
\begin{tabular}{lccl}
Layer & Available data  &  Adopted value   & Remarks \\
      &  $a(\text{U})$ &  $a^{\text{ref}}(\text{U})$      &         \\
\hline
Oceans \& Seawater 
& $3.2\times 10^{-9}$   &  $3.2\times 10^{-9}$  & \cite{hunt83} \\
Sediments  & $1.68\times 10^{-6}$  &  $1.68\times 10^{-6}$  & \cite{plank98} \\
Upper CC   & $ (2.2 \;; 2.4 \; ; 2.5 \; ; 2.8 )\times 10^{-6}$ 
                                  &  $2.5\times 10^{-6}$  & 
Average of~\cite{condie93}, \cite{condie93}, \cite{wedepohl}, \cite{taylor85}\\
Middle CC  & $1.6\times 10^{-6}$  &  $1.6\times 10^{-6}$  & \cite{rudnick95} \\
Lower  CC   & $ (0.20 \; ; 0.28 \; ; 0.93 \; ; 1.1)\times 10^{-6}$ 
                                  &  $0.62\times 10^{-6}$  & 
Average of~\cite{rudnick95}, \cite{taylor85}, \cite{wedepohl}, \cite{shaw86}\\
Oceanic crust  & $0.1\times 10^{-6}$  &  $0.1\times 10^{-6}$  & \cite{taylor85} \\
Upper mantle   & $ (5 \; ; 8)\times 10^{-9}$ 
                                  &  $6.5\times 10^{-9}$  
                                  & Average of~\cite{jochum}, \cite{zartman} \\
Lower mantle  &   &  $13.2\times 10^{-9}$  & From Eq.~(\ref{BSEconstrain}) with\\
              &   &                       
                         & $a_{\text{BSE}}(\text{U})= 2\times 10^{-8}$ \\
Core  &    &   0 &       \\
\end{tabular}
\end{ruledtabular}
\end{table}

\begin{table}
\caption[ccc]
{\label{tab:abundancesTh}
Thorium abundances in Earth's interior.
}
\begin{ruledtabular}
\begin{tabular}{lcccl}
Layer   & Available data & Average            &  Adopted value        & Remarks  \\
        & Th/U         & $\langle \text{Th/U} \rangle $  & $a^{\text{ref}}(\text{Th})$  &  \\

\hline
Oceans \& Seawater     
& 0     & 0      &  0                &   \cite{hunt83} \\
Sediments  & 4.11  & 4.11   & $6.9\times 10^{-6}$  &   \cite{plank98} \\
Upper CC   & $ 3.8 \;; 3.8 \; ; 3.9 \; ; 4.1  $ &  3.9
                                  &  $9.8\times 10^{-6}$  & 
 Average of~\cite{rudnick95}, \cite{condie93}, \cite{condie93}, \cite{wedepohl}   \\
Middle CC  & 3.8   &  3.8   & $6.1\times 10^{-6}$  & \cite{rudnick95} \\
Lower  CC  & $ 3.8 \; ; 6.0 \; ; 7.0 \; ; 7.1 $ & 6  
                                  &  $3.7\times 10^{-6}$  & 
Average of~\cite{taylor85}, \cite{rudnick95}, \cite{shaw86}, \cite{wedepohl} \\
Oceanic crust  & 2.2 & 2.2  &  $0.22\times 10^{-6}$  & \cite{taylor85} \\
Upper mantle   & $ 2.58 \; ; 2.63 \; ; 2.7 \; ; 2.73 $ & 2.66 
                                  &  $17.3\times 10^{-9}$  
   & Average of~\cite{white}, \cite{onions}, \cite{hofmann}, \cite{zartman} \\
Lower mantle  &   &  & $52.0\times 10^{-9}$  &  From Eq.~(\ref{BSEconstrain}) with\\
              &   &                       
                     &    & $a_{\text{BSE}}(\text{Th})= 7.8\times 10^{-8}$ \\
Core  &  &  &   0 &       \\
\end{tabular}
\end{ruledtabular}
\end{table}

\begin{table}
\caption[ddd]
{\label{tab:abundancesK}
Potassium abundances in Earth's interior.
}
\begin{ruledtabular}
\begin{tabular}{lcccl}
Layer   & Available data & Average         &  Adopted value        & Remarks  \\
        & $(\text{K/U})\times 10^{-4}$          
       & $\langle \text{K/U} \rangle  \times 10^{-4}$ &  $a^{\text{ref}}(\text{K})$  &  \\
\hline
Oceans \& Seawater     & $12.5$   &   $12.5$  & 
                          $4.0\times 10^{-4}$            &   \cite{hunt83} \\
Sediments  & $1.0$  &  $1.0$   
                       & $1.7\times 10^{-2}$  &   \cite{plank98} \\
Upper CC   & $ 0.99 \;; 1.0 \; ; 1.03 \; ; 1.10  $ & 
              $1.03$ 
                                &  $2.57\times 10^{-2}$  & 
Average of~\cite{taylor85}, \cite{wedepohl}, \cite{condie93}, \cite{condie93} \\
Middle CC  &  $1.04 $  &   $1.04$    
              & $1.67\times 10^{-2}$  & \cite{rudnick95} \\
Lower  CC  & $ 1 \; ; 1.2 \; ; 1.4 $   & $1.2$
                                  &  $0.72\times 10^{-2}$  & 
Average of~\cite{taylor85}, \cite{shaw86}, \cite{wedepohl} \\
Oceanic crust  & $1.25 $  &  $1.25$  &  
                    $0.125\times 10^{-2}$  & \cite{taylor85} \\
Upper mantle   &    &   
                                  &  $0.78\times 10^{-4}$  
           & From K/U approx. constancy  \\
Lower mantle  &   &  & $1.6\times 10^{-4}$  &  From Eq.~(\ref{BSEconstrain}) with\\
              &   &                       
                     &    & $a_{\text{BSE}}(\text{K})= 2.32\times 10^{-4}$ \\
Core  &  &  &   0 &       \\
\end{tabular}
\end{ruledtabular}
\end{table}

\begin{table}
\caption[fff]
{\label{tab:fluxU}
Uranium: masses, radiogenic heat, and predicted fluxes.
Units are $10^{17}$~kg, TW and $10^6$~cm$^{-2}$~s$^{-1}$, respectively.
The reference values, lower and upper limits are labeled as ref, low,
and high, respectively. Crust summarizes CC and OC; UM (LM) denotes
upper (lower) mantle. 
}
\begin{ruledtabular}
\begin{tabular}{lcccccc}
      &           &         & Himalaya  & Gran Sasso & Kamioka   & Hawaii   \\
      &           &         & $33^{\circ}$~N  $85^{\circ}$~E
                                      & $42^{\circ}$~N  $14^{\circ}$~E
                                      & $36^{\circ}$~N  $137^{\circ}$~E 
                                      & $20^{\circ}$~N  $156^{\circ}$~W                  \\
\hline
      &   $M(\text{U})$ &  $H(\text{U})$ & \multicolumn{4}{c}{$\Phi_{\text{U}}$} \\
\hline
Crust low & 0.206 & 1.960 & 3.337 & 1.913 & 1.594 & 0.218 \\
Crust ref & 0.353 & 3.354 & 5.710 & 3.273 & 2.727 & 0.373 \\
Crust high & 0.413 & 3.920 & 6.674 & 3.826 & 3.187 & 0.436 \\
\hline
UM low & 0.048 & 0.455 & 0.146 & 0.146 & 0.146 & 0.146 \\
UM ref & 0.062 & 0.591 & 0.189 & 0.189 & 0.189 & 0.189 \\
UM high & 0.077 & 0.727 & 0.233 & 0.233 & 0.233 & 0.233 \\
\hline
LM low & 0.147 & 1.399 & 0.288 & 0.288 & 0.288 & 0.288 \\
LM ref & 0.389 & 3.695 & 0.760 & 0.760 & 0.760 & 0.760 \\
LM high & 1.177 & 11.182 & 2.299 & 2.299 & 2.299 & 2.299 \\
\hline
Total low & 0.401 & 3.814 & 3.770 & 2.346 & 2.027 & 0.651 \\
Total ref & 0.804 & 7.639 & 6.659 & 4.222 & 3.676 & 1.322 \\
Total high & 1.666 & 15.828 & 9.206 & 6.358 & 5.720 & 2.968 \\
\end{tabular}
\end{ruledtabular}
\end{table}

\begin{table}
\caption[ggg]
{\label{tab:fluxTh}
Thorium: masses, radiogenic heat, and predicted fluxes.
Units are $10^{17}$~kg, TW and $10^6$~cm$^{-2}$~s$^{-1}$, respectively.
The reference values, lower and upper limits are labeled as ref, low,
and high, respectively. Crust summarizes CC and OC; UM (LM) denotes
upper (lower) mantle. 
}
\begin{ruledtabular}
\begin{tabular}{lcccccc}
      &         &         & Himalaya  & Gran Sasso & Kamioka   & Hawaii   \\
      &         &         & $33^{\circ}$~N  $85^{\circ}$~E
                                      & $42^{\circ}$~N  $14^{\circ}$~E
                                      & $36^{\circ}$~N  $137^{\circ}$~E 
                                      & $20^{\circ}$~N  $156^{\circ}$~W                  \\
\hline
      &   $M(\text{Th})$ &  $H(\text{Th})$ & \multicolumn{4}{c}{$\Phi_{\text{Th}}$} \\
\hline
Crust low & 0.838 & 2.263 & 2.972 & 1.714 & 1.420 & 0.180 \\
Crust ref & 1.450 & 3.915 & 5.141 & 2.964 & 2.456 & 0.311 \\
Crust high & 1.722 & 4.649 & 6.105 & 3.520 & 2.916 & 0.370 \\
\hline
UM low & 0.124 & 0.336 & 0.083 & 0.083 & 0.083 & 0.083 \\
UM ref & 0.166 & 0.447 & 0.111 & 0.111 & 0.111 & 0.111 \\
UM high & 0.207 & 0.558 & 0.138 & 0.138 & 0.138 & 0.138 \\
\hline
LM low & 0.383 & 1.034 & 0.165 & 0.165 & 0.165 & 0.165 \\
LM ref & 1.532 & 4.135 & 0.658 & 0.658 & 0.658 & 0.65 \\
LM high & 4.590 & 12.393 & 1.973 & 1.973 & 1.973 & 1.973 \\
\hline
Total low & 1.346 & 3.633 & 3.220 & 1.961 & 1.668 & 0.428 \\
Total ref & 3.147 & 8.497 & 5.910 & 3.733 & 3.225 & 1.080 \\
Total high & 6.519 & 17.600 & 8.216 & 5.631 & 5.028 & 2.481 \\
\end{tabular}
\end{ruledtabular}
\end{table}

\begin{table}
\caption[hhh]
{\label{tab:fluxK}
Potassium: masses, radiogenic heat, and predicted fluxes.
Units are $10^{21}$~kg, TW, and $10^6$~cm$^{-2}$~s$^{-1}$, respectively.
The reference values, lower and upper limits are labelled as ref, low,
and high, respectively. Crust summarizes CC and OC; UM (LM) denotes
upper (lower) mantle. 
}
\begin{ruledtabular}
\begin{tabular}{lcccccc}
      &         &         & Himalaya  & Gran Sasso & Kamioka   & Hawaii   \\
      &         &         & $33^{\circ}$~N  $85^{\circ}$~E
                                      & $42^{\circ}$~N  $14^{\circ}$~E
                                      & $36^{\circ}$~N  $137^{\circ}$~E 
                                      & $20^{\circ}$~N  $156^{\circ}$~W                  \\
\hline
      &   $M(\text{K})$ &  $H(\text{K})$ & \multicolumn{4}{c}{$\Phi_{\text{K}}$} \\
\hline
Crust low & 0.210 & 0.757 & 12.429 & 7.126 & 5.941 & 0.851 \\
Crust ref & 0.367 & 1.321 & 21.684 & 12.432 & 10.366 & 1.485 \\
Crust high & 0.441 & 1.587 & 26.048 & 14.934 & 12.451 & 1.784 \\
\hline
UM low & 0.057 & 0.207 & 0.634 & 0.634 & 0.634 & 0.634 \\
UM ref & 0.075 & 0.269 & 0.824 & 0.824 & 0.824 & 0.824 \\
UM high & 0.092 & 0.331 & 1.015 & 1.015 & 1.015 & 1.015 \\
\hline
LM low & 0.177 & 0.636 & 1.254 & 1.254 & 1.254 & 1.25 \\
LM ref & 0.471 & 1.697 & 3.343 & 3.343 & 3.343 & 3.34 \\
LM high & 1.344 & 4.838 & 9.534 & 9.534 & 9.534 & 9.534 \\
\hline
Total low & 0.444 & 1.600 & 14.317 & 9.014 & 7.829 & 2.739 \\
Total ref & 0.913 & 3.287 & 25.852 & 16.600 & 14.533 & 5.652 \\
Total high & 1.877 & 6.756 & 36.596 & 25.482 & 23.000 & 12.332 \\
\end{tabular}
\end{ruledtabular}
\end{table}

\begingroup
\squeezetable
\begin{table}
\caption[lll]
{\label{tab:Kami}
Differential produced fluxes: the contributions from the crust at Kamioka.
The distance $R$ is in km, $f_X$ in cm$^{-3}$~s$^{-1}$. 
}
\begin{ruledtabular}
\begin{tabular}{rccc}
$R$   &   $f_{\text{U}}$   &   $f_{\text{Th}}$ &  $f_{\text{K}}$ \\  
\hline
    10 &  1.86E-01 &  1.61E-01 &  6.94E-01\\
    20 &  2.29E-01 &  1.97E-01 &  8.55E-01\\
    30 &  2.01E-01 &  1.75E-01 &  7.57E-01\\
    40 &  1.59E-01 &  1.45E-01 &  6.07E-01\\
    50 &  1.23E-01 &  1.12E-01 &  4.68E-01\\
    60 &  9.86E-02 &  9.04E-02 &  3.76E-01\\
    70 &  8.34E-02 &  7.65E-02 &  3.18E-01\\
    80 &  7.51E-02 &  6.87E-02 &  2.86E-01\\
    90 &  6.62E-02 &  6.06E-02 &  2.52E-01\\
   100 &  5.57E-02 &  5.11E-02 &  2.12E-01\\
   200 &  2.31E-02 &  2.12E-02 &  8.82E-02\\
   300 &  8.15E-03 &  7.39E-03 &  3.12E-02\\
   400 &  5.24E-03 &  4.74E-03 &  2.01E-02\\
   500 &  3.68E-03 &  3.31E-03 &  1.41E-02\\
   600 &  2.61E-03 &  2.35E-03 &  1.00E-02\\
   700 &  2.47E-03 &  2.23E-03 &  9.50E-03\\
   800 &  2.53E-03 &  2.29E-03 &  9.68E-03\\
   900 &  2.94E-03 &  2.67E-03 &  1.13E-02\\
  1000 &  2.88E-03 &  2.61E-03 &  1.10E-02\\
  2000 &  1.32E-03 &  1.20E-03 &  5.06E-03\\
  3000 &  1.08E-03 &  9.72E-04 &  4.11E-03\\
  4000 &  1.05E-03 &  9.51E-04 &  4.01E-03\\
  5000 &  7.44E-04 &  6.75E-04 &  2.84E-03\\
  6000 &  4.88E-04 &  4.40E-04 &  1.86E-03\\
  7000 &  4.28E-04 &  3.86E-04 &  1.64E-03\\
  8000 &  2.99E-04 &  2.69E-04 &  1.14E-03\\
  9000 &  2.53E-04 &  2.27E-04 &  9.67E-04\\
 10000 &  2.19E-04 &  1.98E-04 &  8.41E-04\\
 11000 &  2.16E-04 &  1.96E-04 &  8.28E-04\\
 12000 &  1.40E-04 &  1.24E-04 &  5.35E-04\\
\end{tabular}
\end{ruledtabular}
\end{table}
\endgroup

\begingroup
\squeezetable
\begin{table}
\caption[mmm]
{\label{tab:GrSa}
Differential produced fluxes: the contributions from the crust at Gran Sasso.
The distance $R$ is in km, $f_X$ in cm$^{-3}$~s$^{-1}$. 
}
\begin{ruledtabular}
\begin{tabular}{rccc}
$R$   &   $f_{\text{U}}$   &   $f_{\text{Th}}$ &  $f_{\text{K}}$ \\  
\hline
    10 &  1.48E-01 &  1.29E-01 &  5.48E-01\\
    20 &  2.11E-01 &  1.82E-01 &  7.86E-01\\
    30 &  1.80E-01 &  1.59E-01 &  6.79E-01\\
    40 &  1.40E-01 &  1.28E-01 &  5.34E-01\\
    50 &  1.12E-01 &  1.03E-01 &  4.26E-01\\
    60 &  8.94E-02 &  8.21E-02 &  3.41E-01\\
    70 &  7.66E-02 &  7.04E-02 &  2.92E-01\\
    80 &  6.59E-02 &  6.05E-02 &  2.51E-01\\
    90 &  5.92E-02 &  5.43E-02 &  2.25E-01\\
   100 &  5.22E-02 &  4.79E-02 &  1.99E-01\\
   200 &  2.30E-02 &  2.11E-02 &  8.75E-02\\
   300 &  1.31E-02 &  1.20E-02 &  5.02E-02\\
   400 &  1.14E-02 &  1.04E-02 &  4.34E-02\\
   500 &  9.83E-03 &  8.95E-03 &  3.74E-02\\
   600 &  7.52E-03 &  6.81E-03 &  2.86E-02\\
   700 &  5.98E-03 &  5.43E-03 &  2.27E-02\\
   800 &  5.01E-03 &  4.56E-03 &  1.91E-02\\
   900 &  4.95E-03 &  4.52E-03 &  1.88E-02\\
  1000 &  5.12E-03 &  4.68E-03 &  1.95E-02\\
  2000 &  2.98E-03 &  2.71E-03 &  1.13E-02\\
  3000 &  1.60E-03 &  1.45E-03 &  6.08E-03\\
  4000 &  1.22E-03 &  1.11E-03 &  4.66E-03\\
  5000 &  7.65E-04 &  6.91E-04 &  2.91E-03\\
  6000 &  5.98E-04 &  5.42E-04 &  2.28E-03\\
  7000 &  5.66E-04 &  5.14E-04 &  2.16E-03\\
  8000 &  4.44E-04 &  4.02E-04 &  1.69E-03\\
  9000 &  2.20E-04 &  1.97E-04 &  8.41E-04\\
 10000 &  8.20E-05 &  7.18E-05 &  3.19E-04\\
 11000 &  1.61E-04 &  1.46E-04 &  6.20E-04\\
 12000 &  1.27E-04 &  1.14E-04 &  4.88E-04\\
\end{tabular}
\end{ruledtabular}
\end{table}
\endgroup

\begingroup
\squeezetable
\begin{table}
\caption[nnn]
{\label{tab:mantle}
Differential produced fluxes: the contributions from the mantle.
The distance $R$ is in km, $f_X$ in cm$^{-3}$~s$^{-1}$. 
}
\begin{ruledtabular}
\begin{tabular}{rccc}
$R$   &   $f_{\text{U}}$   &   $f_{\text{Th}}$ &  $f_{\text{K}}$ \\  
\hline
   10 &  0.00E+00 &  0.00E+00 &  0.00E+00\\
    20 &  0.00E+00 &  0.00E+00 &  0.00E+00\\
    30 &  0.00E+00 &  0.00E+00 &  0.00E+00\\
    40 &  0.00E+00 &  0.00E+00 &  0.00E+00\\
    50 &  1.62E-04 &  9.48E-05 &  7.05E-04\\
    60 &  2.91E-04 &  1.70E-04 &  1.27E-03\\
    70 &  3.77E-04 &  2.21E-04 &  1.64E-03\\
    80 &  4.38E-04 &  2.57E-04 &  1.91E-03\\
    90 &  4.84E-04 &  2.83E-04 &  2.11E-03\\
   100 &  5.19E-04 &  3.04E-04 &  2.26E-03\\
   200 &  6.64E-04 &  3.89E-04 &  2.89E-03\\
   300 &  7.08E-04 &  4.14E-04 &  3.08E-03\\
   400 &  7.31E-04 &  4.28E-04 &  3.19E-03\\
   500 &  7.53E-04 &  4.41E-04 &  3.28E-03\\
   600 &  7.71E-04 &  4.51E-04 &  3.36E-03\\
   700 &  8.49E-04 &  5.32E-04 &  3.70E-03\\
   800 &  9.83E-04 &  6.74E-04 &  4.30E-03\\
   900 &  1.09E-03 &  7.83E-04 &  4.75E-03\\
  1000 &  1.17E-03 &  8.70E-04 &  5.12E-03\\
  2000 &  1.49E-03 &  1.22E-03 &  6.56E-03\\
  3000 &  1.48E-03 &  1.24E-03 &  6.51E-03\\
  4000 &  1.11E-03 &  9.27E-04 &  4.88E-03\\
  5000 &  8.88E-04 &  7.41E-04 &  3.90E-03\\
  6000 &  7.40E-04 &  6.17E-04 &  3.25E-03\\
  7000 &  6.34E-04 &  5.29E-04 &  2.79E-03\\
  8000 &  5.54E-04 &  4.63E-04 &  2.44E-03\\
  9000 &  4.93E-04 &  4.11E-04 &  2.17E-03\\
 10000 &  4.24E-04 &  3.53E-04 &  1.86E-03\\
 11000 &  2.35E-04 &  1.91E-04 &  1.03E-03\\
 12000 &  6.10E-05 &  4.11E-05 &  2.66E-04\\
\end{tabular}
\end{ruledtabular}
\end{table}
\endgroup

\begin{table}
\caption[iii]
{\label{tab:error}
Fractional uncertainties of the produced fluxes.
}
\begin{ruledtabular}
\begin{tabular}{lcccc}
$\Delta \Phi / \Phi $ (\%)  &  Himalaya   &  Gran Sasso   & Kamioka  &  Hawaii  \\
\hline
conventional 1 $\sigma$       &  14         &  16           & 17       & 29       \\
\end{tabular}
\end{ruledtabular}
\end{table}

\begin{table}
\caption[kkk]
{\label{tab:events}
Total yields.
$N_{\text{no}}$ is the total number of geoevents (U + Th) in the absence
of oscillations predicted from the reference model for 
$10^{32}$ proton yr  (or in TNU) 
and $\Delta N_{\text{no}}$ is the ``1$\sigma$'' error. 
 $N^{\text{low}}_{\text{no}}$ ($N^{\text{high}}_{\text{no}}$) is the minimal (maximal) 
prediction.
For $\delta m^2 > 4\times 10^{-5}$~eV$^2$ the geoevent
yield is $N=N_{\text{no}} \left[ 1- 0.5 \sin^2(2\theta) \right]$.
}
\begin{ruledtabular}
\begin{tabular}{lrrrr}
Location   & $N_{no}$  & $\Delta N_{no}$ & $N^{low}_{no}$ &  $N^{high}_{no}$\\ 
\hline
Baksan & 91 & 13 & 51 & 131 \\
Hawaii & 22 & 6 & 10 & 49 \\
Himalaya & 112 & 15 & 63 & 154 \\
Homestake & 91 & 13 & 51 & 130 \\
Kamioka & 61 & 10 & 33 & 96 \\
La Palma & 37 & 8 & 19 & 67 \\
LGS & 71 & 11 & 39 & 106 \\
Pyhasalmi & 92 & 13 & 51 & 131 \\
Sudbury & 87 & 13 & 48 & 125 \\
Yucca Mountain & 70 & 11 & 38 & 106 \\ 
\end{tabular}
\end{ruledtabular}
\end{table}

\clearpage

\end{document}